\documentclass[11pt]{article}
\usepackage[a4paper,top=70pt,lmargin=70pt,rmargin=70pt,bottom=70pt]{geometry}

\usepackage{graphicx}

\usepackage[]{cite}

\expandafter\let\csname equation*\endcsname\relax
\expandafter\let\csname endequation*\endcsname\relax

\usepackage{amsmath}
\usepackage[]{amssymb}
\usepackage[]{mathrsfs}
\usepackage[]{eucal}
\usepackage[]{tensor}
\usepackage[]{amsthm}
\usepackage[]{bm}

\usepackage{tikz}
\usepackage{tikz-cd}
\usetikzlibrary{arrows,decorations.pathmorphing,patterns,trees}
\usepackage[americanvoltages,europeanresistors,cuteinductors]{circuitikz}
\newcommand{\btkz}{\begin{tikzpicture}}
\newcommand{\etkz}{\end{tikzpicture}}

\newcommand{\pf}[1]{\mathbf{#1}}
\newcommand{\dd}{\partial}
\newcommand{\hdg}{\star} 
\newcommand{\df}{\mathrm{d}}
\newcommand{\w}{\wedge}
\newcommand{\veps}{\bm{\epsilon}}

\newcommand{\nab}[1]{\nabla_{\!#1}}
\newcommand{\qqd}{\ , \quad}
\newcommand{\bc}{\begin{center}}
\newcommand{\ec}{\end{center}}
\newcommand{\be}{\begin{equation}}
\newcommand{\ee}{\end{equation}}

\newcommand{\F}{\pf{F}}
\newcommand{\Z}{\pf{Z}}
\newcommand{\A}{\pf{A}}

\newcommand{\FF}{\mathcal{F}}
\newcommand{\GG}{\mathcal{G}}
\newcommand{\HH}{\mathcal{H}}
\newcommand{\LL}{\mathscr{L}}

\newcommand{\defeq}{\mathrel{\mathop:}=}
\newcommand{\eqdef}{\mathrel{=\mathop:}}

\usepackage{dsfont}

\newcommand{\rr}{\mathds{R}}

\usepackage[unicode]{hyperref}
\usepackage{xcolor}
\definecolor{pastgreen}{HTML}{669900}
\definecolor{pastblue}{HTML}{336699}
\definecolor{pastred}{HTML}{990000}
\definecolor{linkcol}{HTML}{663333}
\hypersetup{colorlinks,linkcolor={pastblue},citecolor={pastblue},urlcolor={pastblue}}

\theoremstyle{plain} \newtheorem{tm}{Theorem}[section]
\theoremstyle{plain} \newtheorem{lm}[tm]{Lemma}
\theoremstyle{plain} \newtheorem{defn}[tm]{Definition}
\newcommand{\btm}{\begin{tm}}
\newcommand{\etm}{\end{tm}}
\newcommand{\blm}{\begin{lm}}
\newcommand{\elm}{\end{lm}}
\newcommand{\bdefn}{\begin{defn}}
\newcommand{\edefn}{\end{defn}}

\begin{document}

\begin{flushright}
\texttt{ZTF-EP-25-07}

\texttt{RBI-ThPhys-2025-41}
\end{flushright}

\vspace{20pt}

\bc
{\huge Conundrum of regular black holes with nonlinear electromagnetic fields}

\vspace{25pt}

{\Large Ana Bokuli\'c$^a$, Tajron Juri\'c$^b$ and Ivica Smoli\'c$^c$}

\bigskip

e-mails: ana.bokulic@ua.pt, tjuric@irb.hr, ismolic@phy.hr

\medskip

$^a$Departamento de Matem\'atica da Universidade de Aveiro and Centre for Research and Development in Mathematics and Applications (CIDMA), Campus de Santiago, 3810-193 Aveiro, Portugal

\smallskip

$^b$Rudjer Bo\v skovi\'c Institute, Bijeni\v cka cesta 54, HR-10002 Zagreb, Croatia

\smallskip

$^c$Department of Physics, Faculty of Science, University of Zagreb, Bijeni\v cka cesta 32, 10000 Zagreb, Croatia

\smallskip

\ec

\vspace{15pt}

\begin{abstract}
The search for regular black holes with nonlinear electromagnetic fields has sprouted numerous candidates, each exhibiting certain virtues but often accompanied by significant drawbacks. We demonstrate that Komar mass, electric charge and magnetic charge are mutually dependent in regular black holes with nonlinear electromagnetic fields, defined by a Lagrangian which is a function of both electromagnetic invariants, $F_{ab} F^{ab}$ and $F_{ab}{\star F}^{ab}$, regardless of the specific weak field limit of the theory. Also, we generalize one of the key no-go theorems by showing that static, spherically symmetric, electrically charged black holes in a theory respecting the relaxed Maxwellian weak field limit do not admit a bounded Kretschmann scalar. Finally, we address one of the long-standing niche questions, whether regular black hole solutions can exist when both electric and magnetic charges are present, by constructing an exotic family of regular dyonic black holes with nonlinear electromagnetic fields in theories respecting the Maxwellian weak field limit. Mounting evidence suggests that regularizing black holes through simplistic nonlinear extensions of Maxwell's electromagnetism entails a high cost in the form of unorthodox theoretical assumptions.
\end{abstract} 

\vspace{5pt}


\bigskip

\section{Introduction} 

An ever-expanding line of research in theoretical gravitational physics originates from the deceptively simple, yet heavily loaded, question ``Do regular black holes exist?''. Without any additional disclaimer, the answer is resounding \emph{yes}, followed by dramatic \emph{however}. A cheap way to produce innumerable examples is based on a simple trick, sometimes referred to as the ``Synge G-method'' \cite{EG24}: Insert spacetime metric $g_{ab}$, regular in some sense which you consider appropriate (geodesic completeness, bounded curvature invariants, etc.) into gravitational field equation and proclaim ``right-hand side'' to be energy-momentum tensor of some physical matter. A seminal work of this approach in the context of the search for regular black holes is Bardeen's two-page proceedings paper \cite{Bardeen68}, featuring a brief discussion of a modified Schwarzschild metric (a more recent resurrection of the idea has appeared in \cite{SV19}). As was already noted by Bardeen, a basic constraint on the corresponding energy-momentum tensor is given by the various energy conditions \cite{Maeda22}. A far more nontrivial challenge lies in constructing a complete Lagrangian (for all the fields and their interaction), such that the corresponding gravitational-matter field equations admit a solution describing a regular black hole spacetime\footnote{For example, one approach studies regular black holes with de Sitter cores \cite{Dymnikova92,DK15,Vertogradov25}, while more recent proposals consider a signature change at the event horizon \cite{CDBB24,CDBB25} and an infinite tower of higher-curvature Lagrangian corrections in higher-dimensional spacetimes \cite{BCH25}.}. All such candidates may be sorted based on the extent to which they depart from established physical theories and ultimately put under the scrutiny of experiments.

\smallskip

Another key aspect in assessing the physical viability of regular black holes is understanding their potential dynamical formation. Several scenarios have been proposed in which repulsive effects prevent singular collapse. These include mechanisms motivated by quantum gravity, such as effective corrections from the asymptotic safety program \cite{PhysRevLett.132.031401} or phenomenological models \cite{PhysRevD.82.044031}. Regular black holes may also be produced dynamically through purely classical mechanisms, relying on modified gravitational theories \cite{NikodemPopławski2021, PhysRevD.111.104009} or instabilities in AdS backgrounds, which may trigger the creation of de Sitter patches corresponding to the interiors of the regular black holes \cite{PhysRevLett.129.251104}.
Formation mechanisms have also been proposed for a broader class of horizonless compact objects that, based on their observational signatures, could mimic black holes \cite{PhysRevD.80.084023, Herdeiro_2021, Sengo_2024}. From a theoretical perspective, these solitoniclike configurations offer a promising avenue for resolving the singularity problem without departing from the framework of general relativity, often using scalar and Proca fields as their matter content. Boson (see \cite{Shnir:2022lba} for a review) and Proca \cite{BRITO2016291} stars are thought to arise through collapse accompanied by gravitational cooling \cite{PhysRevLett.72.2516, PhysRevD.98.064044}, while the formation of gravastars \cite{gravastar} appears to rely on finely tuned dust sphere collapse and the expansion of a de Sitter core \cite{jampolski2025formationgravastars}.

\smallskip
Over the past quarter of century, a significant effort has been put into investigation of gravitating electromagnetic fields. As we already know that conventional Einstein--Maxwell field equations lead to Kerr--Newmann family of singular black hole solutions, a survey had to be spread over various modified theories. The first breakthrough was made by Ay\'on-Beato and Garc{\'i}a \cite{ABG98}, but it was soon realized that this solution suffers from a serious drawback \cite{Bronnikov00c}, stemming from multivaluedness of the Lagrangian. A couple of years later, the same authors managed \cite{ABG00} to find an electromagnetic Lagrangian which reproduces Bardeen's metric as a magnetically charged black hole. Again, this theory exhibits one technical limitation and one physically unrealistic feature, both of which persist in numerous subsequent proposals. First, the Ay\'on-Beato--Garc{\'i}a electromagnetic Lagrangian has singular behaviour in the weak field limit (in the sense which will be defined more precisely below), making it difficult to connect with the conventional, well-tested Maxwell's electromagnetism. In fact, a series of no-go theorems \cite{Bronnikov00,BSJ22b}  constrain theories that admit regular black hole solutions and satisfy Maxwell's weak field limit. Yet, it is still possible to find theories, via Lagrangian reverse engineering, with a satisfactory weak field limit \cite{BFJS24}, admitting regular, magnetically charged black holes. Second, the Ay\'on-Beato--Garc{\'i}a electromagnetic Lagrangian does not have independent black hole mass and magnetic charge\footnote{This point is somewhat muddled in the original paper \cite{ABG00}, with the mass and the magnetic charge appearing among the Lagrangian parameters. As will be shown in the Appendix \ref{app:Bard}, this technical issue may be resolved with appropriate redefinitions, at the cost of making the magnetic charge a function of mass (or vice versa).}. As far as we are aware of, this feature is common for all the regular black holes found in theories with minimally coupled nonlinear electromagnetic Lagrangians. One of the objectives of the present work is to clarify how universal this phenomenon is.

\smallskip

The paper is organized as follows. In section 2, we give a classification and a brief overview of various gravitating nonlinear electromagnetic theories. Reduction of field equations in the static, spherically symmetric spacetime is presented in section 3. The central result of the paper, the necessity of the mass-charge relation in regular black holes with nonlinear electromagnetic fields, is proven in section 4. A couple of simple generalizations of the no-go theorems discussed in section 5 provide the context for two new families of dyonic regular black hole solutions with nonlinear electromagnetic fields, presented in section 6. Ramifications of the results obtained in the paper are discussed in section 7. In Appendix A we briefly analyse reverse engineering with the rational metric function, while in Appendix B we give several technical remarks about Bardeen's black hole and the corresponding Ay\'on-Beato--Garc{\'i}a electromagnetic Lagrangian.

\smallskip

\emph{Notation and conventions}. We use the ``mostly plus'' metric signature and natural system of units in which $c = G = 4\pi\epsilon_0 = 1$. Partial derivatives of the Lagrangian density $\LL(\FF,\GG)$ are denoted by $\LL_\FF \defeq \dd_\FF \LL$, $\LL_\GG \defeq \dd_\GG \LL$, etc.

\section{Panoramic view of gravitating nonlinear electromagnetic fields} 

Nonlinear electromagnetic (NLE) theories are, in general, defined by Lagrangians which contain electromagnetic tensor $F_{ab}$ and its Hodge dual ${\hdg F}_{ab} \defeq \epsilon_{abcd} F^{cd}/2$. The basic class of NLE Lagrangians are functions of various contractions of $F_{ab}$ and ${\hdg F}_{ab}$ (no additional covariant derivatives). It can be shown \cite{EU14} that such NLE Lagrangians are in fact functions $\LL(\FF,\GG)$ of two electromagnetic invariants,
\be
\FF \defeq F_{ab} F^{ab} \qqd \GG \defeq F_{ab} \, {\hdg F}^{ab} .
\ee
Historically, NLE theories appeared in the 1930s, with two prominent examples: the Born--Infeld Lagrangian \cite{Born34,BI34}, constructed with the specific aim to cure the inconsistencies of Maxwell's electromagnetism associated with the infinite self-energy of the point charges, and the Euler--Heisenberg one-loop QED correction to Maxwell's Lagrangian \cite{HE36}. The ensuing century brought us plethora of proposed NLE theories \cite{Sorokin22,Bronnikov22,AEM21}. Nonlinearities of the electromagnetic field are being tested by the ATLAS Collaboration \cite{ATLAS1,ATLAS2,ATLAS3} and new generations of the high-power lasers at the Extreme Light Infrastructure \cite{ELINP20} (see also \cite{BR13,FB16}).

\smallskip

Any extension of a conventional theory must be examined in the weak field limit, where its predictions can be directly compared to those of the original theory. To this end, we must first define precisely what we mean by the ``weak field limit''. We say that a NLE Lagrangian $\LL(\FF,\GG)$ satisfies

\begin{itemize}
\item[(a)] quasi-Maxwellian weak field (qMWF) limit if partial derivatives $\LL_\FF$ and $\LL_\GG$ are well defined on an open subset of the $\FF$-$\GG$ plane containing interval $\left< -a,0 \right>$ along the $\FF$-axis for some $a > 0$, furthermore $\lim_{\FF\to 0^-} \LL_\FF(\FF,0) = b_1 \ne 0$ and $\lim_{\FF\to 0^-} \LL_\GG(\FF,0) = b_2$;

\item[(b)] Maxwellian weak field (MWF) limit if $\LL(\FF,\GG)$ is class $C^1$ on some neighbourhood of the origin of the $\FF$-$\GG$ plane, such that $\LL_\FF(\FF,\GG) = -\tfrac{1}{4} + O(\HH)$ and $\LL_\GG(\FF,\GG) = O(\HH)$ as $\HH \to 0$, where $\HH \defeq \sqrt{\FF^2 + \GG^2}$ is the ``radial distance'' in the $\FF$-$\GG$ plane.
\end{itemize}

\noindent
Both definitions encapsulate some notion that the electromagnetic Lagrangian is ``Maxwell-like'' for weak fields. The difference is that for the MWF limit we impose conditions on some neighbourhood of the origin of the $\FF$-$\GG$ plane (thus, for a broader family of dyonic solutions), whereas for the qMWF limit we only demand that the Lagrangian has an appropriate limit in the absence of magnetic charge, as the electric field is turned off. The MWF limit is a stricter condition than the qMWF limit: while the former implies the latter, the converse implication does not necessarily follow. For example, the Born--Infeld and Euler--Heisenberg Lagrangians satisfy the MWF limit, while ModMax Lagrangian \cite{Kosyakov20,BLST20} satisfies only qMWF and not the MWF limit. Just to complete the nomenclature, we may denote members of qMWF branch which do not satisfy MWF limit as Lagrangians with the ``directional MWF'' (dMWF) limit. Any NLE Lagrangian $\LL(\FF,\GG)$ that fails even at the qMWF limit will be referred to as ``irregular''. An elementary example of an irregular NLE Lagrangian is the so-called power-Maxwell $\LL(\FF) = C |\FF|^s$, with constant $C \ne 0$ and the exponent $s \ne 1$.

\smallskip

A more complex family of NLE theories are those in which the Lagrangian contains contractions of terms with covariant derivatives, such as $\nab{a} F_{bc}$, $\nab{a} {\hdg F}_{bc}$, $\nab{a} \nab{b} F_{cd}$, etc. Members of this branch are collectively referred to as higher-derivative NLE theories, with the closely related Bopp--Land\'e--Thomas--Podolsky \cite{Bopp40,Lande41,LT41,Podolsky42} and Lee--Wick \cite{LW70} theories serving as the paradigmatic examples (see also \cite{TN14,CdMMPP18,Amorim25,SP25} and references therein). A natural generalization of the MWF limit for higher-derivative NLE Lagrangians begins with identifying a set of independent invariants $(\FF,\GG,\mathcal{I}_3,\dots,\mathcal{I}_n)$ that serve as building blocks. One can then say that a Lagrangian of the form $\LL(\FF,\GG,\mathcal{I}_3,\dots,\mathcal{I}_n)$ satisfies an appropriate weak field limit if it is suitably smooth in a neighborhood of the origin in the formal space $\rr^n$ of invariants (we shall not pursue the technical details of this construction here).

\bc
\begin{figure}[ht]
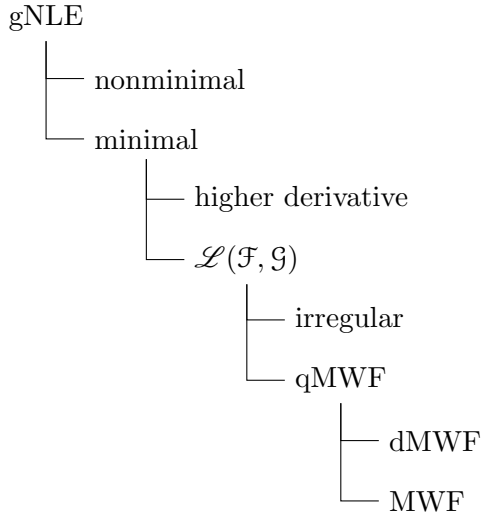

\centering
\btkz
[every node/.append style = {anchor = west},
 grow via three points={one child at (0.5,-0.8) and two children at (0.5,-0.8) and (0.5,-1.6)},
 edge from parent path={(\tikzparentnode\tikzparentanchor) |-(\tikzchildnode\tikzchildanchor)}
] 
\node{gNLE}
 child {node {nonminimal}}
 child {node {minimal}
  child {node {higher derivative}}
  child {node {$\LL(\FF,\GG)$}
   child {node {irregular}} 
   child {node {qMWF}
    child {node {dMWF}}
    child {node {MWF}}}
 }
};
\etkz

\caption{``gNLE tree'': Visual representation of the classification of gravitating NLE theories described in the main text.}\label{fig:tree}
\end{figure}
\ec

We now turn to electromagnetic fields interacting with gravity. Gravitating nonlinear electromagnetic (gNLE) theories may be broadly categorized into those with minimal and those with nonminimal coupling between the gravitational and the electromagnetic fields. In a minimally coupled theory, we simply have a sum of a gravitational Lagrangian and an NLE Lagrangian, a member of some family described above. Nonminimally coupled terms appear in quantum field effective actions \cite{DH80}, and are far more challenging to deal with \cite{BL05}. They could be classified according to the number of pairs of contracted indices (see also \cite{CLN24,CN25}), in which one belongs either to $F_{ab}$, Hodge dual ${\hdg F}_{ab}$ or their covariant derivatives, while the other belongs to Riemann tensor $R_{abcd}$ or its covariant derivatives (for example, one could denote $\LL(\FF,\GG) R$ as type-0 term, $F^{ac} \tensor{F}{^b_c} R_{ab}$ as type-2 term, $F^{ab} F^{cd} R_{abcd}$ as type-4 term, etc.). Astrophysical observational constraints on some of these terms was recently analysed in \cite{HNS25}. Overall, the described classification of gravitating NLE theories may be visually represented with a tree diagram \ref{fig:tree}. 

\smallskip

In the present work, we analyse the $\LL(\FF,\GG)$-branch of gNLE theories, with the standard Einstein--Hilbert gravitational action. Using the auxiliary 2-form $\Z \defeq -4(\LL_\FF\,\F + \LL_\GG\,{\hdg \F})$, the corresponding source-free NLE Maxwell's equations can be written concisely as
\be\label{eq:NLE}
\df\F = 0 \qqd \df{\hdg\Z = 0} .
\ee
Furthermore, Einstein gravitational field equation reads
\be\label{eq:EinstNLE}
R_{ab} - \frac{1}{2} \, R g_{ab} = 8\pi T_{ab} ,
\ee
with the NLE energy-momentum tensor
\be
T_{ab} = -4\LL_\FF \widetilde{T}_{ab} + \frac{1}{4} \, T g_{ab} ,
\ee
written with Maxwell's electromagnetic energy-momentum tensor
\be
\widetilde{T}_{ab} \defeq \frac{1}{4\pi} \left( F_{ac} \tensor{F}{_b^c} - \frac{1}{4} \, \FF \, g_{ab} \right) ,
\ee
and the trace of the energy-momentum tensor,
\be
T \defeq \frac{1}{\pi} \, (\LL - \LL_\FF \, \FF - \LL_\GG \, \GG) .
\ee

\section{Static, spherically symmetric case} 

In this paper, our main focus is on the static, spherically symmetric, asymptotically flat solutions of the Einstein--NLE field equations. More precisely, spacetime consists of a smooth 4-manifold $M$, with an asymptotically flat Lorentzian metric $g_{ab}$ and the electromagnetic 2-form $F_{ab}$. Furthermore, metric $g_{ab}$ admits a hypersurface orthogonal Killing vector field $k^a$, timelike in the asymptotic region, and the group of isometries contains a subgroup isomorphic to $SO(3)$, whose orbits are 2-spheres which foliate hypersurfaces to which $k^a$ is orthogonal. On each of the orbit 2-spheres we may define spherical coordinates $(\theta,\varphi)$, such that they are constant along integral curves of a smooth congruence of geodesics orthogonal to the group orbits. Furthermore, the radial coordinate is introduced as the function $r=(A/4\pi)^{1/2}$ of the area $A$ of the orbit 2-spheres. Leaving aside ``wormhole'' and ``horn'' geometries (see discussion in \cite{Bronnikov00}), we assume that $\nabla_a r \ne 0$ everywhere in the spacetime and radial coordinate takes values on the interval $r \in \left[ 0,\infty \right>$. Finally, Killing parameter $t$, defined via $k^a \nab{a} t = 1$, may be used to complete the coordinate system into canonical $(t,r,\theta,\varphi)$, given that $k^a$ and $\nabla^a r$ are not collinear. As is well known, the latter case occurs on Killing horizons (on which $k^a$ is lightlike), thus the constructed coordinate system is valid only on domains between adjacent Killing horizons, while for the extension across the horizon one must resort to new coordinates tailored along convenient geodesics (such as Eddington--Finkelstein and Kruskal coordinates). Note that the causal character of vectors $(k^a,\nabla^a r)$ may switch from (timelike, spacelike) in one region to (spacelike, timelike) in other region, so that the names ``time coordinate'' and ``radial coordinate'' should be taken with a grain of salt.

\smallskip

Field equations imply \cite{BSJ22b,BJS24} that the product $g_{tt} g_{rr}$ of metric components is constant, hence without loss of generality we may assume that the spacetime metric has the following form 
\be\label{eq:m}
\df s^2 = -f(r) \, \df t^2 + \frac{\df r^2}{f(r)} + r^2 \big( \df\theta^2 + \sin^2\theta \, \df\varphi^2 \big)
\ee
in the canonical coordinate system. In order to avoid pathological examples, we assume that the function $f(r)$ has a finite number of zeros (i.e.~spacetime has a finite number of Killing horizons). Asymptotic flatness and the definition of Komar mass $M$ imply that the function $f$ satisfies
\be
f(r) = 1 - \frac{2M}{r} + O(r^{-2}) \hspace{7pt} \textrm{as} \ r \to \infty , 
\ee
that is
\be
\lim_{r\to\infty} \big( r(f(r)-1) \big) = -2M .
\ee
Regularity of the metric is defined in a sense that the associated Kretschmann scalar $K \defeq R_{abcd} R^{abcd}$ is a bounded continuous real function on the interval $\left[ 0,\infty \right>$. In particular, we say that a metric has a ``regular center'' if $\lim_{r\to 0^+} K$ is well defined and finite. Kretschmann scalar $K$ for the metric (\ref{eq:m}) is a sum of three nonnegative terms,
\be
K(r) = \big( f''(r) \big)^2 + \frac{4}{r^2} \, \big( f'(r) \big)^2 + \frac{4}{r^4} \, \big( f(r)-1 \big)^2 .
\ee
Thus, regular center requires that all three functions, $f''(r)$, $f'(r)/r$ and $(f(r)-1)/r^2$, remain bounded as $r \to 0^+$.

\smallskip

Electromagnetic field, 2-form $F_{ab}$, may be decomposed with respect to the Killing vector field $k^a$ via two 1-forms, electric $E_a \defeq k^b F_{ab}$ and magnetic $B_a \defeq k^b {\hdg F}_{ba}$. If we assume that the electromagnetic field $F_{ab}$ inherits\footnote{Broader discussion about symmetry inheritance of NLE fields is given in \cite{BGS17}, while isometry-consistent block-diagonalizability of the metric for gravitational field equations with NLE field is rigorously analysed in \cite{BS23}.} all spacetime symmetries, according to Lemma 2.1 in \cite{BJS24} it can be written as
\be
\F = -E_r(r) \, \df t \w \df r + B_r(r) r^2 \sin\theta \, \df\theta \w \df\varphi .
\ee
Corresponding electromagnetic invariants are $\FF = 2(B_r^2 - E_r^2)$ and $\GG = 4E_r B_r$. Using the definitions of electric charge $Q$ and magnetic charge $P$,
\be
Q \defeq \frac{1}{4\pi} \, \oint_\infty {\hdg\Z} \qqd P \defeq \frac{1}{4\pi} \, \oint_\infty \F ,
\ee
the source-free NLE Maxwell’s equations are reduced \cite{DARG10,BSJ22b} to the system
\be\label{eq:NLEMax}
\LL_\FF E_r - \LL_\GG B_r = -\frac{Q}{4r^2} \qqd B_r = \frac{P}{r^2} .
\ee
Furthermore, Einstein’s gravitational field equations are reduced to
\begin{align}
\frac{(r(f(r) - 1))'}{2r^2} & = \LL + 4(\LL_\FF E_r - \LL_\GG B_r) E_r , \\
\frac{f''(r)}{4} + \frac{f'(r)}{2r} & = \LL - 4(\LL_\FF B_r + \LL_\GG E_r) B_r .
\end{align}
Alternatively, using NLE Maxwell's equations (\ref{eq:NLEMax}), the first equation may be simplified to
\be\label{eq:Einst1alt}
\frac{\big( r(f(r) - 1) \big)'}{2r^2} = \LL - \frac{Q}{r^2} \, E_r .
\ee
Note that a bounded Kretschmann scalar implies that left-hand sides of both gravitational field equations remain bounded as $r \to 0^+$. As was shown by two theorems in \cite{BSJ22a}, strictly stationary, regular solutions of the Einstein-NLE field equations, with the energy-momentum tensor respecting the dominant energy condition, have either a trivial or stealth electromagnetic field. Thus, our primary concern is with the regular black hole solutions.

\section{Mass and charges entangled} 

We shall first investigate the question of independence of mass and charges for regular black hole solutions. Formally, Lagrangian is a function of both electromagnetic invariants and finite set of Lagrangian parameters, $\LL(\FF,\GG;\lambda_1,\dots,\lambda_n)$, but Lagrangian parameters $(\lambda_1,\dots,\lambda_n)$ are fixed in the variational procedure for the field equations and subsequent analysis of the solutions, so we may set them aside. For clarity, we first consider the magnetic case before extending the analysis to the more general dyonic case.

\smallskip

A purely magnetic black hole is defined as one with zero electric charge, $Q = 0$, nonvanishing magnetic charge, $P \ne 0$, and an identically vanishing electric field, $E_r = 0$. Consequently, we have $\FF = 2P^2/r^4$, $\GG = 0$ and the field equation (\ref{eq:Einst1alt}) may be written as
\be
\big( r(f(r) - 1) \big)' = 2r^2 \LL(2P^2/r^4,0) .
\ee
Now, using asymptotic properties of the metric function $f$, and relying on the assumption that $f$ and its derivative $f'$ are at least continuous across horizons, we may evaluate an elementary integral
\be\label{eq:theint}
\int_\infty^r \big( s(f(s) - 1) \big)' \, \df s = r \big( f(r) - 1 \big) + 2M
\ee
for any $r > 0$. On the other hand, with the substitution $u = 2P^2/s^4$ we have
\be\label{eq:Lagint}
-2 \int_r^\infty s^2 \LL(2P^2/s^4,0) \, \df s = -\frac{|P|^{\frac{3}{2}}}{2^{\frac{1}{4}}} \int_0^{\tfrac{2P^2}{r^4}} \!\!\LL(u,0) u^{-\frac{7}{4}} \, \df u \eqdef |P|^{\frac{3}{2}} h(P/r^2) .
\ee
The regularity of the center implies that $\lim_{r\to 0^+} (f(r)-1) = 0$, ensuring that the integral on the left-hand side of equation (\ref{eq:theint}) is well-defined and finite\footnote{Alternatively, given that $\LL(u,0) = O(u^\alpha)$ as $u \to 0^+$, with $\alpha > 3/4$, and $\LL(u,0) = O(u^\beta)$ as $u \to +\infty$, with $\beta < 3/4$, ensures convergence of the integral $I$.}, and so is the integral
\be
I \defeq \lim_{r\to 0^+} h(P/r^2) = -2^{-\frac{1}{4}} \int_0^\infty \LL(u,0) u^{-\frac{7}{4}} \, \df u .
\ee
We may rewrite integrals (\ref{eq:theint}) and (\ref{eq:Lagint}) in a convenient form, with the explicit mass and charge variables,
\be
\frac{f(r;M,P)-1}{r^2} = \frac{-2M + |P|^{\frac{3}{2}} h(P/r^2)}{r^3} .
\ee
Again, regularity of the center demands $\lim_{r\to 0^+} \big( {-2M} + |P|^{\frac{3}{2}} h(P/r^2) \big) = 0$, that is
\be\label{eq:MP32}
M = \frac{I}{2} \, |P|^{\frac{3}{2}} .
\ee
In other words, to ensure that the Kretschmann scalar is bounded, the Komar mass $M$ and the magnetic charge $P$ must be constrained by the equation above. A special, pathological case occurs if $I = 0$, implying that we need $M = 0$ for regularization. Relation (\ref{eq:MP32}) should not come as a surprise, as it was already present in Ay\'on-Beato--Garc{\'i}a's NLE interpretation of the Bardeen's black hole (see comments in Appendix \ref{app:Bard}) and it has appeared in more recent reverse engineered NLE Lagrangians \cite{BFJS24} (see comments in Appendix \ref{app:ratio}). 

\smallskip

This conclusion can be extended to a more general, dyonic case with electric and magnetic charge. We begin by assuming that mass and charges $(M,Q,P)$ are independent variables and show how the condition of the regular black hole center leads to a contradiction. Formally, we may say that the magnetic field is a function of radial coordinate and magnetic charge, $B_r = B_r(r;P)$, whereas for the electric field we must \emph{a priori} assume a more general form $E_r = E_r(r;M,Q,P)$. Taking the partial derivatives of the first NLE Maxwell equation with respect to $Q$ and $M$, we obtain conditions
\begin{align}
\left( - 4\LL_{\FF\FF} E_r^2 + 8\LL_{\FF\GG} \, \frac{P E_r}{r^2} - 4\LL_{\GG\GG} \, \frac{P^2}{r^4} + \LL_\FF \right) \frac{\dd E_r}{\dd Q} & = -\frac{1}{4 r^2} , \\
\left( - 4\LL_{\FF\FF} E_r^2 + 8\LL_{\FF\GG} \, \frac{P E_r}{r^2} - 4\LL_{\GG\GG} \, \frac{P^2}{r^4} + \LL_\FF \right) \frac{\dd E_r}{\dd M} & = 0 .
\end{align}
The only consistent option is $\dd E_r/\dd M = 0$. Consequently, neither $\FF$ nor $\GG$ depends on the mass $M$, and the field equation (\ref{eq:Einst1alt}) may be written in the form
\be
\big( r(f(r) - 1) \big)' = 2r^2 \LL(\FF(r;Q,P),\GG(r;Q,P)) - 2QE_r(r;Q,P) .
\ee
Relying on the same strategy as above, after the integration, we get a relation
\be
\frac{f(r;M,Q,P)-1}{r^2} = \frac{-2M + \mathcal{I}(r;Q,P)}{r^3} ,
\ee
where
\be
\mathcal{I}(r;Q,P) \defeq -2 \int_r^\infty \big( s^2 \LL(\FF(s;Q,P),\GG(s;Q,P)) - QE_r(s;Q,P) \big) \, \df s .
\ee
Finally, regularity of the center implies $M = \lim_{r\to 0^+} \mathcal{I}(r;Q,P)/2$, revealing that mass and charges in general cannot remain independent (in contradiction with the initial assumption). In other words, regular black holes with NLE fields reside within a special subset of the parameter space, the physical viability of which will be examined in section 7.

\section{Generalized constraints} 

Let us turn to another kind of constraint on regular black holes with NLE fields. Back in 2001 Bronnikov \cite{Bronnikov00} proved that a NLE Lagrangian $\LL(\FF)$ with the MWF limit does not admit electrically charged, static, spherically symmetric black holes with bounded Ricci scalar $R$ and ``Ricci squared'' scalar $S \defeq R_{ab} R^{ab}$. This result was later generalized \cite{BSJ22b} to a larger class of $\LL(\FF,\GG)$ Lagrangians. Using the terminology from section 2, in combination with the recent results from \cite{DMS25}, it is straightforward to extend and strengthen the conclusion even further, in the form of the following theorem.

\btm\label{tm:no-go}
Suppose that the spacetime is a static, spherically symmetric solution of the Einstein-NLE field equations (\ref{eq:NLE}) and (\ref{eq:EinstNLE}) with the NLE Lagrangian $\LL(\FF,\GG)$ obeying the qMWF limit and charges $Q \ne 0$ and $P = 0$. Then the Kretschmann scalar $K$ cannot remain bounded as $r \to 0^+$.  
\etm

\noindent
\emph{Proof}. The main part of the proof is identical to the proof of Theorem 1 in \cite{BSJ22b}, reaching the conclusion that, under given assumptions, $R$ and $S$ cannot both remain bounded as $r \to 0^+$. Then, due to inequalities $R^2 \le 4S$ and $2S \le 3K$ proven in \cite{DMS25} (see recent generalizations in \cite{SKN25}), it follows that Kretschmann scalar $K$ is unbounded as well. \qed

\medskip

This generalization immediately applies to theories respecting qMWF and not necessarily the MWF limit, such as ModMax theory. Another perspective may be added here, as ModMax theory belongs to the larger class of NLE theories with vanishing trace of energy-momentum tensor, $T=0$. Einstein's gravitational field equation with the cosmological constant $\Lambda$ (included here for the sake of generality) implies for any member of such a theory that the Ricci scalar is constant, so that for static, spherically symmetric spacetimes considered here, we have 
\be
4\Lambda = R = f''(r) + \frac{4}{r} \, f'(r) + \frac{2}{r^2} \, (f(r)-1) .
\ee
This equation has solution of the form $f(r) = 1 + C_1/r + C_2/r^2 + (\Lambda/3) r^2$, with constants $C_1$ and $C_2$, for which the Kretschmann scalar is not bounded as $r \to 0^+$,
\be
K(r) = \frac{4}{r^8} \, \big( 14 C_2^2 + 12 C_1 C_2 r + 3 C_1^2 r^2 \big) + \frac{8\Lambda^2}{3} ,
\ee
unless $C_1 = C_2 = 0$.

\smallskip

Analogous no-go results \cite{BSJ22b} hold for dyonic solutions of several restricted families of NLE Lagrangians, such as $\LL(\FF)$ and quadratic $\LL(\FF,\GG) = -\FF/4 + a\FF^2 + b\FF\GG + c\GG^2$ with constants $(a,b,c)$. However, once we look past such special families of theories, generalizations become more challenging across the rest of the charge $(Q,P)$ parameter space. As will be shown in the next section, the origin of these obstacles are families of NLE theories and dyonic solutions with regular black holes that, to our knowledge, have not been previously noticed.

\section{Dyonic exotica} 

Backbone of numerous regular black hole solutions with NLE fields is the Lagrangian reverse engineering, reconstruction of the NLE Lagrangian from a given spacetime metric. The original procedure \cite{Bronnikov00,FW16,Bronnikov17c,TSA18} (see also more recent generalization \cite{JJ25} for NLE fields coupled to scalar fields), tailored for magnetically charged black holes and $\FF$-dependent Lagrangians $\LL(\FF)$, was recently \cite{BFJS24} refined to allow adjusting of higher order terms in the weak field limit of the NLE Lagrangian. A dyonic solution-generating scheme with $\LL(\FF)$ Lagrangians was proposed in \cite{Bronnikov17}. Nevertheless, due to the aforementioned no-go theorem, it is already clear that this approach cannot yield a regular black hole, provided the NLE Lagrangian respects the qMWF limit. The remaining open question is the generalization of the reverse engineering procedure for dyonic black holes and NLE Lagrangians depending on both electromagnetic invariants, $\LL(\FF,\GG)$. We shall demonstrate how to ``recycle'' the original trick and produce two novel families of NLE Lagrangians with the sought properties.

\subsection{First family} 

Useful examples are sometimes obtained through an ansatz that, at first glance, may appear to be an \emph{ad hoc} choice. Motivated by the simple idea of preserving the form of the electric and magnetic fields as closely as possible, a process of trial and error led us to a family of NLE Lagrangians,
\be\label{eq:Lagpsi}
\LL(\FF,\GG) = \psi\left( \frac{\FF + \alpha\GG}{1+\alpha^2} \right) + \beta \GG ,
\ee
where $\alpha$ and $\beta$ are constants, and $\psi$ is an auxiliary function which will be defined below. As a ``special ingredient'', Lagrangian (\ref{eq:Lagpsi}) contains the boundary $\GG$-term (in the action we have $\GG \, \bm{\epsilon} = -2\,\df (\A \w \F)$ with the volume 4-form $\veps$ and the gauge 1-form $\A$), which does not affect the field equations, but it contributes to the charges, akin to the so-called Witten effect \cite{Witten79,WittenEffect}. Magnetic field is already fixed, $B_r = P/r^2$, while for the electric field we choose the ansatz\footnote{This form of the electric field is reminiscent of the ModMax case \cite{FAGMLM21,BH25a,BH25b}, in which $E_r = Q e^{-\gamma}/r^2 = (Q/P) e^{-\gamma} B_r$, although the ModMax Lagrangian does not belong to the family (\ref{eq:Lagpsi}).} $E_r = \alpha B_r$. Consequently, we have $\FF = 2(1-\alpha^2) P^2/r^4$ and $\GG = 4\alpha P^2/r^4$, so that $\FF + \alpha\GG = 2(1+\alpha^2) P^2/r^4$.

\smallskip

Now, given that the charges are related via $Q = 4\beta P$, source-free NLE Maxwell's equations are satisfied. Furthermore, the first gravitational field equation (\ref{eq:Einst1alt}) is reduced to
\be
\frac{\big( r(f(r)-1) \big)'}{2r^2} = \psi\big( 2P^2/r^4 \big) .
\ee
The strategy of Lagrangian reverse engineering turns this equation upside down: starting from a function $f$, chosen such that (along with other desirable properties) it has a bounded Kretschmann invariant as $r \to 0^+$, we have a ``recipe'' for the Lagrangian function $\psi$,
\be\label{eq:RevEng}
\psi(x) = \frac{\big( r(f(r)-1) \big)'}{2r^2} \, \Bigg|_* \ ,
\ee
where the asterisk ``$*$'' denotes the substitution $r = (2P^2/x)^{1/4}$. Finally, using the first gravitational field equation and its derivative with respect to the radial coordinate $r$, it is straightforward to check that the second gravitational field equation is satisfied.

\smallskip

Finally, with an appropriate choice of the relation between $\alpha$ and $\beta$ we can ensure that the MWF limit is satisfied. Given that the function $\psi$ is at least $C^1$ on some neighbourhood of the origin, we can use the Taylor series with remainder,
\be
\psi\left( \frac{\FF + \alpha\GG}{1+\alpha^2} \right) = \psi(0) + \psi'(0) \, \frac{\FF + \alpha\GG}{1+\alpha^2} + O\big( (\FF + \alpha\GG)^2 \big) .
\ee
Constant $\psi(0)$ is irrelevant, as it represents the constant term in the Lagrangian. Part of the weak field limit may be fixed if $\psi'(0) = -(1+\alpha^2)/4$, while the $\GG$-term in the Lagrangian may be effectively ``removed'' at least in the weak field limit with the choice $\beta = -\alpha\psi'(0)/(1+\alpha^2) = \alpha/4$. More concretely, if the metric function has an asymptotic form
\be\label{eq:fasymptalpha}
f(r) = 1 - \frac{2M}{r} + \frac{(1+\alpha^2) P^2}{r^2} + O(r^{-3})
\ee
as $r \to \infty$, then
\be
\psi'(x) = \frac{\df r}{\df x} \, \frac{\df}{\df r} \frac{\big( r(f(r)-1) \big)'}{2r^2} \, \Bigg|_* = -\frac{1+\alpha^2}{4} + O(x^{1/4})
\ee
as $x \to 0^+$. Consequently, with these choices we have $Q=4\beta P = \alpha P$ and $(1+\alpha^2) P^2 = Q^2 + P^2$ (metric (\ref{eq:fasymptalpha}) asymptotically coincides with the dyonic Reissner--Nordstr\"om). The remaining technical issue of the procedure is the systematic replacement of the mass and charges with Lagrangian parameters. An example, based on a rational metric function $f(r)$, is analysed in the Appendix \ref{app:ratio}. We note in passing that any of these NLE Lagrangians (satisfying the MWF limit) are counterexamples to the no-go theorem 1.5 in \cite{Bronnikov22}, which cannot hold as stated and presumably relies on additional assumptions.

\smallskip

As a brief digression, it is interesting to look more closely at the special case $\alpha = 0 = \beta$, in which the NLE Maxwell's equations are reduced to the equation $\psi'(\FF) E_r(r) = 0$. Trivial electric field, $E_r = 0$, is a part of the family described above, and is essentially the well-known reverse engineering with magnetically charged regular black holes. The other branch stems from the nontrivial condition $\psi'(\FF) = 0$. Suppose that we use ansatz (\ref{eq:RevEng}) for a given convenient function $f(r)$, leading to the function $\psi$ such that its derivative $\psi'$ has at least one negative zero, e.g.~$\psi'(-C) = 0$ for some $C \ge 0$. Then, with the choice of the electric field $E_r = \sqrt{B_r^2 + C/2}$, we have constant $\FF = -C$ and $\psi'(\FF) = 0$. However, this new branch of solutions is physically unacceptable as $E_r$ does not vanish as $r \to \infty$, unless $C = 0$. Even worse, gravitational field equation imply $(r(f-1))'/(2r^2) = \psi = f''/4 + f'/(2r)$, so that $f(r) = -1 + C_1 r^2 + C_2/r$, with constants $C_1$ and $C_2$, whose corresponding Kretschmann scalar is unbounded.

\subsection{Second family} 

Another family of regular black hole solutions is built from the Lagrangian ansatz with separated $\FF$ part and $\GG$ part \cite{BFJS24},
\be
\LL(\FF,\GG) = \mathcal{J}(\FF) + \mathcal{K}(\GG) ,
\ee
such that $\mathcal{K}(0) = 0$ and $\kappa \defeq \mathcal{K}_\GG(0) \ne 0$. Here we choose a trivial electric field, $E_r = 0$, and impose a constraint on charges $Q = 4\kappa P$, which implies that NLE Maxwell's equations are satisfied. The rest of the field equations
\begin{align}
\frac{f'(r)}{2r} + \frac{f(r)-1}{2r^2} & = \mathcal{J}(\FF) , \\
\frac{f''(r)}{4} + \frac{f'(r)}{2r} & = \mathcal{J}(\FF) - \frac{4P^2}{r^4} \, \mathcal{J}'(\FF) ,
\end{align}
are equivalent to the well-known magnetic case with the $\FF$-dependent Lagrangian. Thus, using the usual Lagrangian reverse engineering (again, see Appendix \ref{app:ratio}) we may choose a metric function $f(r)$ such that the total Lagrangian $\LL(\FF,\GG)$ satisfies the qMWF limit (an obstacle for the full MWF limit is nonvanishing parameter $\kappa \ne 0$, which is essential for the presence of the electric charge in this construction). Note that, due to Theorem 3 in \cite{BSJ22b}, $\mathcal{J}(\FF)$ cannot be simply Maxwell's $-\FF/4$. In conclusion, we have a dyonic regular black hole, alas with the vanishing electric field. Overlap between two introduced families of Lagrangians and solutions occurs when $\mathcal{K}(\GG) = \kappa\GG$, $\alpha = 0$, and $\beta = \kappa$.

\section{Discussion} 

One could argue that the search for regular black holes with NLE fields represents a relatively conservative approach. Rather than introducing new fields or modifying the gravitational Lagrangian, the focus has remained on well-tested physics grounded in extensions of Maxwell’s electromagnetism. The results presented here allow for a critical evaluation of this entire endeavor and help to illuminate various theoretical impasses.

\smallskip

Starting with the NLE Lagrangians of the form $\LL(\FF,\GG)$, previous results \cite{Bronnikov00,BSJ22b}, as well as novel Theorem \ref{tm:no-go}, have demonstrated that there are no static, spherically symmetric, electrically charged, regular black holes with $\LL(\FF,\GG)$ Lagrangian respecting the qMWF limit. In other words, in order to obtain regular black hole solutions, one must admit at least one of the two features: irregular Lagrangian or magnetic charges. The former choice represents a theoretical challenge, as we need to make some sense of the weak field limit and connection with the canonical Maxwell's electromagnetism. On the other hand, the introduction of magnetic charges, at least initially, does not seem like a radical idea. It is true that at the moment of writing this paper, there is still no experimental evidence of magnetic charges in our universe. Nevertheless, there are some theoretical suggestions that they might exist and it is possible, in principle, that we need just a minuscule amount of magnetic charge in order to regularize the black holes.

\smallskip

Unfortunately, here we encounter a much serious obstacle. Analysis in section 4 reveals that a bounded Kretschmann scalar $K(r)$ necessitates that Komar mass $M$, electric charge $Q$ and magnetic charge $P$ cannot remain independent black hole parameters. The relation between mass and charges is not completely unheard of in theoretical physics, with examples such as dyonic BPS states \cite{SW94}. However, given that we strive to give an argument for the regularity of \emph{generic} static black holes, we have to reach for additional ``theoretical gymnastics''. First of all, one needs to fine-tune the mass-charge relation in order to make it consistent with numerous astrophysical observations of macroscopic objects (we are not aware of any physical mechanism that would naturally lead to the desired black hole mass-charge constraint). Even worse, the implementation of the universal mass-charge relation at the level of known particles is bound to fail. Let us assume that the hypothetical magnetic charge is conserved and antiparticles carry opposite magnetic charge with respect to their particle counterparts. Now, both the $\pi^0$ meson and $Z$ boson may decay into particle-antiparticle pairs, hence we may deduce that $P_{\pi^0} = P_Z = 0$. Furthermore, $\pi^0$ and $Z$ are also electrically neutral, so that universal relation of the form $M = \mu(Q,P)$ would imply that $M_{\pi^0} = M_Z = \mu(0,0)$, which is certainly not true! Of course, none of these arguments do not completely eliminate the possibility of such a NLE theory, as one could extend the hypothetical mass-charge relation into $M = \mu(Q,P,\dots)$ with some additional variables (quantum numbers). Nonetheless, the obtained constraints inevitably push the construction of such a theory into the realm of, to put it mildly, unorthodox theoretical assumptions.

\smallskip

On top of all the aforementioned constraints, any proposed theory must also confront the delicate question of stability. Namely, even if a given NLE theory admits regular black hole solutions, their physical relevance is significantly undermined if those solutions are unstable. While earlier studies \cite{MS03,TSSA18,NYS20} indicate that certain NLE theories possess regions of parameter space where the black hole exteriors are free from instabilities, more recent work \cite{DeFT25}, focused on $\FF$-dependent Lagrangians, highlights the presence of universal instabilities in black hole interiors.

\smallskip

Returning once more to the panoramic perspective, the theories in other branches of the gNLE tree possess field equations that are technically more challenging. Nevertheless, it was shown \cite{CM21} that at least some of them admit electrically charged, regular black hole solutions, without the necessity for the mass-charge relation present in the $\LL(\FF,\GG)$-branch. A major challenge moving forward is to develop a systematic classification of gNLE theories, guided by their weak-field limits, the types of black hole regularizations they allow, and their stability properties.

\section*{Acknowledgements}
The research was supported by the Croatian Science Foundation Project No. IP-2025-02-8625, \emph{Quantum aspects of gravity}. The research of A.~B. is supported by CIDMA under the Portuguese Foundation for
Science and Technology (FCT, https://ror.org/00snfqn58) Multi-Annual
Financing Program for R\&D Units, grants UID/4106/2025 and UID/PRR/
4106/2025, as well as the projects: Horizon Europe staff exchange (SE) programme HORIZON-MSCA2021-SE-01 Grant No. NewFunFiCO-101086251;  2022.04560.PTDC (\url{https://doi.org/10.54499/2022.04560.PTDC}) and 2024.05617.CERN (\url{https://doi.org/10.54499/2024.05617.CERN}). 


\appendix 

\section{Regular black holes with rational metric function}\label{app:ratio} 

Very simple, yet sufficiently rich metric ansatz \cite{BFJS24} is based on the rational function
\be
f(r) = 1 - \frac{A(r)}{B(r)} ,
\ee
where $A$ and $B$ are polynomials,
\be
A(r) = a_m r^m + \cdots + a_{m+k} r^{m+k} \qqd B(r) = b_n r^n + \cdots + b_{n+\ell} r^{n+\ell} ,
\ee
with nonnegative integers $m,n,k,\ell \ge 0$. In order to avoid trivial cases, we assume that $n+\ell \ge 1$, $a_{m+k} \ne 0$ and $b_{n+\ell} \ne 0$. Furthermore, asymptotic flatness and regularity of the metric are implemented as follows:
\begin{itemize}
\item[(a)] Asymptotic form $f(r) = 1 + O(r^{-1})$ as $r \to \infty$ holds given that $m+k+1 = n+\ell$;
\item[(b)] Kretschmann scalar $K(r)$ is bounded as $r \to 0^+$ given that $A/B = O(r^2)$, $(A/B)' = O(r)$ and $(A/B)'' = O(1)$ as $r \to 0^+$, that is $m \ge n+2$.
\end{itemize}
Key information about the weak field limit of the reverse engineered Lagrangian hides in the expansion
\begin{align}
\frac{\big( r(f(r)-1) \big)'}{2r^2} & = \frac{a_{m+k}}{2b_{n+\ell}} \, (p-1) r^{-(p+2)} + \nonumber \\
 & + \frac{p a_{m+k-1} b_{n+\ell} + (p-2) a_{m+k} b_{n+\ell-1}}{2b_{n+\ell}^2} \, r^{-(p+3)} + O(r^{-(p+4)})
\end{align}
as $r \to \infty$, with the abbreviation $p \defeq n-m + \ell-k$. Taking into account condition (a), it follows that $p=1$ and 
\be
\frac{\big( r(f(r)-1) \big)'}{2r^2} = \frac{a_{m+k-1} b_{n+\ell} - a_{m+k} b_{n+\ell-1}}{2b_{n+\ell}^2} \, r^{-4} + O(r^{-5}) .
\ee
Since the leading term is of the order $O(r^{-4})$, the MWF limit may be obtained with the appropriate choice of the polynomial coefficients. More concretely, if we compare further details of the expansion
\be
f(r) = 1 - \frac{a_{m+k}}{b_{n+\ell}} \, r^{-1} + \frac{a_{m+k} b_{n+\ell-1} - a_{m+k-1} b_{n+\ell}}{b_{n+\ell}^2} \, r^{-2} + O(r^{-3})
\ee
as $r \to \infty$, with the ansatz asymptotics (\ref{eq:fasymptalpha}), we are lead to the conditions
\be\label{eq:abcond}
a_{m+k} = 2M b_{n+\ell} \qqd a_{m+k} b_{n+\ell-1} - a_{m+k-1} b_{n+\ell} = (1+\alpha^2) P^2 b_{n+\ell}^2 \, .
\ee
The first technical challenge is the systematic replacement of the mass and charge parameters with the Lagrangian parameters in the reverse engineered Lagrangian. Instead of a general algorithm we shall illustrate this procedure with one concrete example. Let us take $A(r) = a_2 r^2$ and $B(r) = b_0 + b_1 r + b_2 r^2 + b_3 r^3$. Conditions (\ref{eq:abcond}) boil down to $a_2 = 2Mb_3$ and $b_2 = (1+\alpha^2) P^2 b_3/(2M)$. In addition, we know from the earlier discussion that we have to assume that $M = \gamma |P|^{3/2}$, with some constant $\gamma$, in order to consistently assure a bounded Kretschmann scalar. Finally, we shall apply substitutions $b_i = \lambda_i/|P|^{(i+3)/2}$, where $\lambda_i$ are Lagrangian parameters. All together, we get
\be\label{eq:psiexample}
\psi(x) = -\gamma\lambda_3 \, \frac{\omega\lambda_3 x + 2^{5/4} \lambda_1 x^{5/4} + 3\lambda_0 x^{3/2}}{(2^{3/4} \lambda_3 + \omega\lambda_3 x^{1/4} + 2^{1/4} \lambda_1 x^{1/2} + \lambda_0 x^{3/4})^2} ,
\ee
with the abbreviation $\omega \defeq (1+\alpha^2)/(2^{1/2} \gamma)$. As can be easily verified, the obtained function $\psi$ satisfies $\psi(x) = -(1+\alpha^2) x/4 + O(x^{5/4})$ as $x \to 0^+$. One of the basic assumptions $b_3 \ne 0$ implies $\lambda_3 \ne 0$, hence the denominator in (\ref{eq:psiexample}) is nonzero at least on some interval $\left[ 0,x_0 \right>$. As a consequence, function $\psi$, which is initially formally defined on the interval $\left[ 0,x_0 \right>$, can be extended into a $C^1$ function across the origin $x=0$ via $\psi(-x) = -\psi(x)$ for $x \in \left< 0,x_0 \right>$.

\section{Comments about Bardeen's black hole}\label{app:Bard} 

Bardeen's \cite{Bardeen68} static, spherically symmetric spacetime metric (\ref{eq:m}) has metric function
\be
f_{\mathrm{Bard}}(r) = 1 - \frac{2Mr^2}{(r^2 + r_0^2)^{3/2}} ,
\ee
with Komar mass $M$ and parameter $r_0 > 0$. The corresponding Kretschmann scalar,
\be
K_{\mathrm{Bard}}(r) = \frac{12M^2 (4r^8 - 12 r_0^2 r^6 + 47 r_0^4 r^4 - 4 r_0^6 r^2 + 8r_0^8)}{(r^2 + r_0^2)^7} ,
\ee
is bounded as $r \to 0^+$. The reverse engineering procedure allows us to interpret Bardeen's metric as a solution (magnetically charged black hole) of the Einstein-NLE field equations with the NLE Lagrangian
\be
\LL(\FF) = \frac{\big( r(f_{\mathrm{Bard}}(r)-1) \big)'}{2r^2} \, \Bigg|_{r = (2P^2/\FF)^{1/4}} = -3Mr_0^2 \left( \frac{\sqrt{\FF}}{|P|\sqrt{2} + r_0^2 \sqrt{\FF}} \right)^{\!\frac{5}{2}} .
\ee
Ay\'on-Beato and Garc\'ia \cite{ABG00} correctly conclude that $r^2 B_r$ is a constant (magnetic charge $P$), but then jump to the conclusion that $r_0 = P$ (they use notation $g \defeq r_0$), without any proper justification. In order to get rid of the solution parameters (mass and magnetic charge) from the Lagrangian, it is more convenient to choose $r_0^2 = \lambda_1 |P|$, with some Lagrangian parameter $\lambda_1$, leading to 
\be
\LL(\FF) = -3\lambda_1 M |P|^{-\frac{3}{2}} \left( \frac{\sqrt{\FF}}{\sqrt{2} + \lambda_1 \sqrt{\FF}} \right)^{\!\frac{5}{2}} .
\ee
Then, as a second step, we may introduce another Lagrangian parameter $\lambda_2 \defeq -3\lambda_1 M |P|^{-3/2}$, so that the Lagrangian is reduced to
\be\label{eq:LrevBard}
\LL(\FF) = \lambda_2 \left( \frac{\sqrt{\FF}}{\sqrt{2} + \lambda_1 \sqrt{\FF}} \right)^{\!\frac{5}{2}} ,
\ee
while mass and magnetic charge are related via $M = -\lambda_2 |P|^{3/2}/(3\lambda_1)$, as predicted earlier. The reconstructed Lagrangian (\ref{eq:LrevBard}) is defined only for $\FF \ge 0$, and unfortunately, there is no sensible way to extend it across the origin $\FF=0$ to a function that complies with the qMWF limit since $\lim_{\FF\to 0^+} \LL_\FF(\FF) = 0$.

\bibliographystyle{amsalpha}
\bibliography{NLEcon}

\end{document}